\documentclass[12pt]{article}
\usepackage{amssymb}
\usepackage{amsmath,amsfonts,feynmp}
\usepackage{tipa}
\usepackage{stmaryrd}

\makeatletter\@addtoreset{equation}{section}\makeatother

\def\be{\begin{equation}}
\def\ee{\end{equation}}
\def\bea{\begin{eqnarray}}
\def\eea{\end{eqnarray}}

\makeatletter\@addtoreset{equation}{section}\makeatother

\renewcommand{\title}[1]{\vbox{\center\LARGE{#1}}\vspace{5mm}}
\renewcommand{\author}[1]{\vbox{\center#1}\vspace{5mm}}
\newcommand{\address}[1]{\vbox{\center\em#1}}

\begin{document}

\unitlength = .8mm

\begin{titlepage}
\begin{center}
\hfill \\
\hfill \\
\vskip 1cm

\title{ Note on New Massive Gravity in $AdS_3$ }

{Yan Liu\footnote{Email: liuyan@itp.ac.cn}}, {Ya-Wen
Sun\footnote{Email: sunyw@itp.ac.cn}}

\address{ Key Laboratory of Frontiers in Theoretical Physics£¬
\\ Institute of Theoretical Physics, Chinese Academy of Sciences}

\end{center}

\vskip 1cm

\abstract{ In this note we study the properties of linearized
gravitational excitations in the new massive gravity theory in
asymptotically $AdS_3$ spacetime and find that there is also a
critical point for the mass parameter at which massive gravitons
become massless as in topological massive gravity in $AdS_3$.
However, at this critical point in the new massive gravity the
energy of all branches of highest weight gravitons vanish and the
central charges also vanish within the Brown-Henneaux boundary
conditions. The new massive gravity in asymptotically $AdS_3$
spacetime seems to be trivial at this critical point under the
Brown-Henneaux boundary conditions if the Brown-Henneaux boundary
conditions can be consistent with this theory. At this point, the
boundary conditions of log gravity may be preferred.}

\vfill

\end{titlepage}


\section{Introduction}

The AdS/CFT correspondence
\cite{Maldacena:1997re,{Gubser:1998bc},{Witten:1998qj},{Aharony:1999ti}}
has given rise to increasing interest in the study of gravity theory
in asymptotically $AdS_3$ spacetime. In three dimensions, the
Riemann tensor can be fully determined by the Ricci tensor and there
are no locally propagating degrees of freedom in pure gravity
theory. However, with a negative cosmological constant black hole
solutions \cite{Banados:1992wn} can be found which obey the laws of
black hole thermodynamics and posses a nonzero entropy. For pure
gravity in $AdS_3$ the CFT dual of this quantum gravity has been
identified in \cite{{Witten:1988hc},Witten:2007kt}. More discussions
on $AdS_3/CFT_2$ correspondence can be found in
\cite{{Manschot:2007zb},{Gaiotto:2007xh},{Gaberdiel:2007ve},
{Avramis:2007gx},{Yin:2007gv},{Yin:2007at},{Maloney:2007ud},
{Manschot:2007ha},{Gaiotto:2008jt},{Giombi:2008vd},Kraus:2006wn}.

To develop further understanding to quantum gravity on
asymptotically $AdS_3$ spacetime, a deformation of pure Einstein
gravity theory named topological massive gravity
\cite{{Deser:1981wh},{Deser:1982vy}} with a negative cosmological
constant has been studied \cite{Li:2008dq}. In this topological
massive gravity with a negative cosmological constant, it was
conjectured that there exists a chiral point at which the theory
becomes chiral with only right-moving modes. Later it was proved
that it is indeed chiral at the critical point
\cite{Strominger:2008dp,{Carlip:2008qh}}. More discussions on this
theory can be found in
\cite{{Carlip:2008jk},{Hotta:2008yq},{Grumiller:2008qz},
{Li:2008yz},{Park:2008yy},{Grumiller:2008pr},{Carlip:2008eq},
{Gibbons:2008vi},{Anninos:2008fx},{Giribet:2008bw},
{Compere:2008cv},{Grumiller:2008es},{Garbarz:2008qn},{Blagojevic:2008bn},
{Henneaux:2009pw},Maloney:2009ck}.

Recently in \cite{Bergshoeff:2009hq} a new kind of massive gravity
has been discovered in three dimensions. In this new massive
gravity, higher derivative terms are added to the Einstein Hilbert
action and unlike in topological massive gravity, parity is
preserved in this new massive gravity. This new massive gravity is
equivalent to the Pauli-Fierz action for a massive spin-2 field at
the linearized level in asymptotically Minkowski spacetime. In
\cite{Nakasone:2009bn}, the unitarity of this new massive gravity
was examined and this new massive gravity is generalized to higher
dimensions, but the unitarity is violated in higher dimensions.
Warped AdS black hole solutions for this new massive gravity with a
negative cosmological constant have been found in
\cite{Clement:2009gq}.

In this paper we study the behavior of the linearized gravitational
excitations of this new massive gravity with a negative cosmological
constant in the background of $AdS_3$ spacetime with Brown-Henneaux
boundary conditions. We find that similar to gravitons in
topological massive gravity in $AdS_3$ background, we also have a
critical value of the mass parameter at which massive gravitons
become massless. However, in this new massive gravity, the energy of
both highest weight massless and massive gravitons are all zero and
the central charge is also zero at the critical point. As long as
the Brown-Henneaux boundary conditions can be consistent with this
theory, the theory looks trivial at the critical point. But we need
further studies on the consistency of the Brown-Henneaux boundary
condition and the conserved charges associated with the symmetry.

In the remainder of this paper, we will first review the new massive
gravity with a negative cosmological constant and write out the
central charge in Sec.2. In Sec.3 we will calculate the behavior of
the linearized gravitational excitations around $AdS_3$. In Sec.4 we
show that to have an $AdS_3$ vacuum we can only have a critical
value of the mass parameter but at this special point the theory
seems to be trivial. Sec.5 is devoted to conclusions and
discussions.

\section{The New Massive Gravity Theory}
In this section we review the basics of the new massive gravity with
a negative cosmological constant. The action of the new massive
gravity theory can be written as\footnote{ We take the metric
signature(-,+,+) and follow the notation and conventions of MTW
\cite{Misner:1974qy}. We assume $m^2>0$ and $G$ is the three
dimensional Newton constant which is positive
here.}\cite{Bergshoeff:2009hq}

 \be I=\frac{1}{16\pi
G}\int d^3x \sqrt{-g} \bigg[R-2\lambda-\frac{1}{m^2}K\bigg], \ee
where \be K=R^{\mu\nu}R_{\mu\nu}-\frac{3}{8}R^2,\ee $m$ is the mass
parameter of this massive gravity and $\lambda$ is a constant which
is different from the cosmological constant. The Einstein equation
of motion of this action is \be G_{\mu\nu}+\lambda
g_{\mu\nu}-\frac{1}{2m^2}K_{\mu\nu}=0 \label{eom}\ee where
 \be K_{\mu\nu}=-\frac{1}{2}\nabla^2 R
g_{\mu\nu}-\frac{1}{2}\nabla_\mu\nabla_\nu R+2\nabla^2 R_{\mu\nu}
+4R_{\mu \alpha \nu \beta}R^{\alpha
\beta}-\frac{3}{2}RR_{\mu\nu}-R_{\alpha\beta}R^{\alpha\beta}g_{\mu\nu}+\frac{3}{8}R^2g_{\mu\nu}.
\ee One special feature of this choice of $K$ is that $g^{\mu\nu}
K_{\mu\nu}=K.$

 To have an asymptotically
$AdS_3$ solution, we have to introduce a non-zero $\lambda$. For an
$AdS_3$ solution
\begin{equation}\label{AdS3metric}
ds^2=\bar{g}_{\mu\nu}dx^\mu dx^\nu=\ell^2(-
\cosh^2{\rho}d\tau^2+\sinh^2{\rho}d\phi^2+d\rho^2)
\end{equation}
with an AdS radius $\ell$ which is related to the cosmological
constant $\Lambda$ by
\begin{equation}
\ell^{-2}=-\Lambda,
\end{equation}
the Riemann tensor, Ricci tensor and Ricci scalar of the $AdS_3$ are

\begin{equation}
\bar{R}_{\mu\alpha\nu\beta} = \Lambda (\bar{g}_{\mu\nu}
\bar{g}_{\alpha\beta} - \bar{g}_{\mu\beta} \bar{g}_{\alpha\nu})  ,
\quad \bar{R}_{\mu\nu} = 2 \Lambda \bar{g}_{\mu\nu}  , \quad \bar{R}
= 6 \Lambda  ,
\end{equation}and $K_{\mu\nu}=-1/2\Lambda^2g_{\mu\nu}$. Thus the
$\lambda$ in the action should be related to the cosmological
constant $\Lambda$ and the mass parameter by
 \be
m^2=\frac{\Lambda^2}{4(-\lambda+\Lambda)}.\ee Note here for a given
$\lambda<0$, there can be both AdS and de-Sitter solutions to this
action. We only focus on the AdS solution.

The metric has an isometry group $SL(2,R)_L\times SL(2,R)_R$. The
$SL(2,R)_L$ generators are \cite{Li:2008dq}
\begin{eqnarray}
L_0 &= &i\partial_u, \\ L_{-1} &= &  i e^{-iu} \left[ { \cosh 2 \rho
\over \sinh 2 \rho } \partial_u - { 1 \over \sinh 2 \rho}
\partial_v +  { i \over 2} \partial_\rho \right] ~, \\ L_{1} &= &  i
e^{iu} \left[ { \cosh 2 \rho \over \sinh 2 \rho } \partial_u - { 1
\over \sinh 2 \rho} \partial_v -  { i \over 2} \partial_\rho
\right],\end{eqnarray} where $u \equiv \tau + \phi, v \equiv
\tau-\phi$. The $SL(2,R)_R$ generators $\{\bar{L}_0,
\bar{L}_{\pm1}\}$ are given by the above expressions with $u
\leftrightarrow v$. The central charge of this gravity theory in
asymptotically $AdS_3$ spacetime can be calculated using the formula
\cite{{Saida:1999ec},{Kraus:2005vz},{Park:2006zw},Kraus:2006wn}
\begin{equation}
c=\frac{\ell}{2G}\bar{g}_{\mu\nu}\frac{\partial \mathcal
{L}}{\partial R_{\mu\nu}},
\end{equation}
where $\mathcal{L}$ is the Lagrangian density, to be \be c=
\frac{3\ell}{2G}\bigg(1-\frac{1}{2m^2\ell^2}\bigg).\ee Unlike
topological massive gravity, because we have no Chern-Simons term
here the left moving and right moving central charges are equal and
we have
\begin{equation}
c_{L}=c_{R}=\frac{3\ell}{2G}\bigg(1-\frac{1}{2m^2\ell^2}\bigg).
\end{equation}
We can calculate the entropy of BTZ black holes using Cardy formula
as in \cite{Li:2008dq} and we find that the entropy obtained from
the Cardy formula is just the same as the one obtained in
\cite{Clement:2009gq} using Wald's formula. In order to get a
non-negative central charge we need to set $m^2\ell^2\geq1/2$. Here
we see that $m^2<0$ can also give positive central charge and
entropy. However, in the next section we will see that $m^2<0$ is
not allowed. The mass of the BTZ black hole is also non-negative in
this parameter region as can be found in \cite{Clement:2009gq}.

\section{Gravitons in $AdS_3$}
In this section we analyze the behavior of linearized gravitational
excitations on the background $AdS_3$ spacetime in this new massive
gravity theory.
\subsection{ The equation of motion for graviton}
We first give the equation of motion for gravitons in this
subsection. By expanding
\begin{equation}
g_{\mu\nu} = \bar{g}_{\mu\nu} + h_{\mu\nu}
\end{equation}
with $h_{\mu\nu}$ small, we have the following physical quantities
to the leading order
\be\Gamma^{\lambda
(1)}_{\mu\nu}=\frac{1}{2}\bar{g}^{\lambda\alpha}(\bar{\nabla}_{\mu}h_{\alpha\nu}+
\bar{\nabla}_\nu h_{\mu\alpha}-\bar{\nabla}_\alpha h_{\mu\nu}),\ee
\be R^{\lambda(1)}_{~\rho\nu\mu} = \frac{1}{2}
\bar{g}^{\lambda\alpha}(\bar{\nabla}_\nu\bar{\nabla}_{\mu}h_{\alpha\rho}+
\bar{\nabla}_\nu\bar{\nabla}_\rho
h_{\mu\alpha}-\bar{\nabla}_\nu\bar{\nabla}_\alpha
h_{\mu\rho})-(\mu\leftrightarrow \nu),\ee
\begin{eqnarray}
R_{\mu\nu}^{(1)}& =& \frac{1}{2} (- \bar{\nabla}^2  {h}_{\mu\nu} -
\bar{\nabla}_{\mu}  \bar{\nabla}_{\nu}  h + \bar{\nabla}^{\sigma}
\bar{\nabla}_{\nu}
 h_{\sigma\mu} + \bar{\nabla}^{\sigma}  \bar{\nabla}_{\mu}  h_{\sigma\nu}),\\
R^{(1)} &\equiv& (R_{\mu\nu}  g^{\mu\nu})^{(1)} = - \bar{\nabla}^2 h
+ \bar{\nabla}_{\mu} \bar{\nabla}_{\nu}  h^{\mu\nu} - 2 \Lambda h,
\end{eqnarray}
\begin{eqnarray}
(\nabla_\mu \nabla_\nu R)^{(1)}& =& \bar{\nabla}_\mu \bar{\nabla}_\nu R^{(1)},\\
(\nabla_\alpha \nabla_\beta R_{\mu\nu})^{(1)} &=&
\bar{\nabla}_\alpha \bar{\nabla}_\beta R_{\mu\nu}^{(1)}-2\Lambda
\bar{\nabla}_\alpha \bar{\nabla}_\beta h_{\mu\nu}.
\end{eqnarray}

Substituting these quantities to the equation of motion (\ref{eom}),
we have the equation of motion for graviton $h_{\mu\nu}$ to be
\be G^{(1)}_{\mu\nu}+\lambda
h_{\mu\nu}-\frac{1}{2m^2}K^{(1)}_{\mu\nu}=0 \label{eom forh},\ee
where \bea
G^{(1)}_{\mu\nu}&=& R^{(1)}_{\mu\nu}-\frac{1}{2}\bar{g}_{\mu\nu}R^{(1)}-3\Lambda h_{\mu\nu},  \\
K^{(1)}_{\mu\nu} &=& -\frac{1}{2}\bar{\nabla^2} R^{(1)}
\bar{g}_{\mu\nu}-\frac{1}{2}\bar{\nabla}_\mu\bar{\nabla}_\nu
R^{(1)}+2\bar{\nabla^2} R^{(1)}_{\mu\nu}-4\Lambda
\bar{\nabla}^2h_{\mu\nu}\nonumber\\&&-5\Lambda R^{(1)}_{\mu\nu}
+\frac{3}{2}\Lambda R^{(1)}\bar{g}_{\mu\nu}+\frac{19}{2}\Lambda^2
h_{\mu\nu}.\eea Taking the trace of the equation of motion by
multiplying $\bar {g}^{\mu\nu}$ on both sides, we obtain
 \be R^{(1)}=0.\ee Then we fix the gauge as it was done in \cite{Li:2008dq}. We define
$\tilde{h}_{\mu\nu} \equiv h_{\mu\nu}-\bar{g}_{\mu\nu}h$, which
gives $\tilde{h}=-2h$ and \be
h_{\mu\nu}=\tilde{h}_{\mu\nu}-\frac{1}{2}\bar{g}_{\mu\nu}\tilde{h},\ee
\begin{equation}
R^{(1)}=\bar{\nabla}_{\mu}\bar{\nabla}_\nu\tilde{h}^{\mu\nu}+\Lambda
\tilde{h}=0.
\end{equation}
Thus the gauge
\begin{equation}
\bar{\nabla}_\mu\tilde{h}^{\mu\nu}=0
\end{equation} together with the linearized equation of motion
implies tracelessness of $h_{\mu\nu}$: $\tilde{h}=-2h=0$. This gauge
is equivalent to the harmonic plus traceless gauge
$\bar{\nabla}_{\mu}h^{\mu\nu}=h=0$.

Noting that
\begin{equation}\label{cmt}
[\bar{\nabla}_\sigma,\bar{\nabla}_\mu]h^\sigma_{\nu}=3\Lambda
h_{\mu\nu}-\Lambda h g_{\mu\nu}\end{equation} and imposing the gauge
condition, we obtain \be
R^{(1)}_{\mu\nu}=\frac{1}{2}(-\bar{\nabla}^2h_{\mu\nu}+6\Lambda
h_{\mu\nu}).\ee By using $R^{(1)}=0$ and gauge fixing conditions
$\bar{\nabla}_{\mu}h^{\mu\nu}=h=0$, we reach the following
simplified equation for graviton $h_{\mu\nu}$: \be
G^{(1)}_{\mu\nu}+\lambda h_{\mu\nu}-\frac{1}{2m^2}K^{(1)}_{\mu\nu}=0
\label{eom forh},\ee with \bea G^{(1)}_{\mu\nu}&=&
R^{(1)}_{\mu\nu}-3\Lambda h_{\mu\nu}
=-\frac{1}{2}\bar{\nabla}^2h_{\mu\nu} , \\
K^{(1)}_{\mu\nu}
&=&-\bar{\nabla}^2\bar{\nabla}^2h_{\mu\nu}+\frac{9}{2}\Lambda
\bar{\nabla}^2h_{\mu\nu}-\frac{11}{2}\Lambda^2 h_{\mu\nu}.\eea Thus
the equation of motion for graviton could be factorized as \be
\Big[(\bar{\nabla}^2-(m^2+\frac{5\Lambda}{2})\Big]
\Big[\bar{\nabla}^2-2\Lambda\Big]h_{\mu\nu}=0.
\label{eomforgraviton}\ee From the equation of motion above, we can
easily see that there are two branches of solutions. The first is
\be\Big[\bar{\nabla}^2-2\Lambda\Big]h_{\mu\nu}=0,\ee which
corresponds to the modes of the left and right moving massless
gravitons, and the second one is
\be\Big[(\bar{\nabla}^2-(m^2+\frac{5\Lambda}{2})\Big]h_{\mu\nu}=0,\ee
which corresponds to massive sectors of gravitons. To have a
non-negative mass, we also need to have $m^2\ell^2\geq1/2$ which
coincides with the condition needed to ensure a non-negative central
charge. To give a positive mass of the gravitons, $m^2<0$ is not
allowed. It can be seen that when $m^2\ell^2= 1/2$, the massive
sector of gravitons becomes massless. It is very similar to chiral
gravity at the critical point. In the next subsection we will solve
the equation of motion of gravitons.

\subsection{Solutions of gravitons}

Using the expressions of the generators of the isometry group
$SL(2,R)_L\times SL(2,R)_R$ of $AdS_3$, we could easily simplify the
equation of motion for gravitons using \be\bar{\nabla}^2h_{\mu\nu}=
-[{2\over\ell^ 2}(L^2+\bar{L}^2)+{6\over\ell^ 2}]h_{\mu\nu}\ee to be
\be [-\frac{2}{\ell^ 2}(L^2+\bar{L}^2)-\frac{7}{2\ell^
2}-m^2][{1\over \ell^ 2}(L^2+\bar{L}^2)+\frac{2}{\ell^
2}]h_{\mu\nu}=0\ . \ee We follow \cite{Li:2008dq} to classify the
solutions of (\ref{eomforgraviton}) using the $SL(2,R)_L\times
SL(2,R)_R$ algebra. Considering highest weight states with weight
$(h,\bar{h})$ and using
$L^2|\psi_{\mu\nu}\rangle=-h(h-1)|\psi_{\mu\nu}\rangle$, we obtain
\be [2h(h-1)+2\bar{h}(\bar{h}-1)-\frac{7}{2}-m^2\ell^ 2]
[h(h-1)+\bar{h}(\bar{h}-1)-2]=0, \quad h-\bar{h}=\pm 2\ee for the
highest weight states. There are two branches of solutions for this
equation. The first one is the massless one
$h(h-1)+\bar{h}(\bar{h}-1)-2=0$, which gives \be
h={3\pm1\over2},\,\bar{h}={-1\pm1\over2}\quad \hbox{or}\quad
h={-1\pm1\over2},\,\bar{h}={3\pm1\over2}. \ee The solutions with the
lower sign will blow up at infinity as argued in \cite{Li:2008dq},
so we will only keep the upper ones corresponding to weights $(2,0)$
and $(0,2)$. We will refer to these as left and right-moving
massless gravitons.

The second branch has
$2h(h-1)+2\bar{h}(\bar{h}-1)-\frac{7}{2}-m^2\ell^ 2=0$, which gives
\be h=\frac{6\pm \sqrt{2+4m^2\ell^2}}{4},\,\bar{h}=\frac{-2\pm
\sqrt{2+4m^2\ell^2}}{4}\ee
\\or \be h=\frac{-2\pm
\sqrt{2+4m^2\ell^2}}{4},\,\bar{h}=\frac{6\pm
\sqrt{2+4m^2\ell^2}}{4}.\ee

The solutions with the lower sign will also blow up at the infinity,
so we will only keep the ones with the upper sign which correspond
to weights $(\frac{6+ \sqrt{2+4m^2\ell^2}}{4},\frac{-2+
\sqrt{2+4m^2\ell^2}}{4})$ and $(\frac{-2+
\sqrt{2+4m^2\ell^2}}{4},\frac{6+ \sqrt{2+4m^2\ell^2}}{4})$. We refer
to these modes as massive gravitons. Here we should also have
$m^2\ell^2\geq 1/2$ in order that there is no blow up at the
infinity. The point $m^2\ell^2=1/2$ is very interesting, because the
massive gravitons become massless at this point\footnote{At this
critical point, there can be other solutions \cite{Grumiller:2008qz}
of the graviton which, however, do not obey the Brown-Henneaux
boundary conditions
\cite{{Grumiller:2008es},{Henneaux:2009pw},Maloney:2009}. We would
like to thank Wei Song for pointing this out to us.}.

\section{Positivity of Energy}

In this section, we follow the method of \cite{Li:2008dq} to
calculate the energy of the linearized gravitons in $AdS_3$
background. We assume Brown-Henneaux boundary conditions
\cite{Brown:1986nw} in the following calculation. However, the
consistency of this kind of condition with the new massive gravity
still needs to be further confirmed.

The fluctuation $h_{\mu\nu}$ can be decomposed as
 \begin{eqnarray}h_{\mu\nu}&=&h^{M}_{\mu\nu}+h^{m}_{\mu\nu},\end{eqnarray}
and here we use "M" to denote "massive" modes and "m" to denote
 "massless" modes.

Up to total derivatives, the quadratic action of $h_{\mu\nu}$ can be
written as \bea S_2&=&-\frac{1}{32\pi G}\int d^3x \sqrt{-\bar{g}}
h^{\mu\nu}(G^{(1)}_{\mu\nu}+\lambda
h_{\mu\nu}-\frac{1}{2m^2}K^{(1)}_{\mu\nu})\\
&=&-\frac{1}{32\pi G}\int d^3x
 \sqrt{-\bar{g}}\{\frac{1}{2m^2}\bar{\nabla}^2h^{\mu\nu}\bar{\nabla}^2
 h_{\mu\nu}+(\frac{1}{2}-
 \frac{9}{4m^2\ell^2})\bar{\nabla}^\lambda
h^{\mu\nu}\bar{\nabla}_\lambda
h_{\mu\nu}+(\frac{11}{4m^4\ell^2}+\lambda)h^{\mu\nu}h_{\mu\nu}\}.\nonumber\eea
The momentum conjugate to $h_{\mu\nu}$ is \be
\Pi^{(1)\mu\nu}={\sqrt{-\bar{g}}\over32\pi G}\{
\frac{1}{m^2}\bar{\nabla}^0\bar{\nabla}^2h^{\mu\nu}
-(1-\frac{9}{2m^2\ell^2})\bar{\nabla}^0h^{\mu\nu}\},\ee and using
the equations of motion we can have \bea
\Pi_{m}^{(1)\mu\nu}&=&-{\sqrt{-\bar{g}}\over32\pi
G}(1-\frac{5}{2m^2\ell^2})\bar{\nabla}^0h_m^{\mu\nu},\\
\Pi_{M}^{(1)\mu\nu}&=&{\sqrt{-\bar{g}}\over32\pi
G}(\frac{2}{m^2\ell^2}) \bar{\nabla}^0h_M^{\mu\nu}.\eea Because we
have up to four time derivatives in the Lagrangian, using the
Ostrogradsky method \cite{Li:2008dq,{Buchbinder:1992pe}} we should
also introduce $K_{\mu\nu}\equiv\bar{\nabla}_0h_{\mu\nu}$ as a
canonical variable, whose conjugate momentum is \be
\Pi^{(2)\mu\nu}=-\frac{\sqrt{-\bar{g}}}{32\pi
G}\frac{\bar{g}^{00}}{m^2}\bar{\nabla}^2h^{\mu\nu},\ee and again
using equations of motion we can have \bea
\Pi^{(2)\mu\nu}_m&=&\frac{\sqrt{-\bar{g}}}{32\pi G}
\frac{2\bar{g}^{00}}{m^2\ell^2}h_m^{\mu\nu},\\
\Pi^{(2)\mu\nu}_M&=&-\frac{\sqrt{-\bar{g}}\bar{g}^{00}}{32\pi
G}(1-\frac{5}{2m^2\ell^2})h_M^{\mu\nu}.\eea The Hamiltonian is then
expressed using these variables to be \be H=\int
d^2x\bigl(\dot{h}_{\mu\nu}\Pi^{(1)\mu\nu}+\dot{K}_{\mu\nu}\Pi^{(2)\mu\nu}
-\mathcal{L}\bigr).\ee

After substituting the equations of motion of the highest weight
states, we can have the energies as follows \bea \label{energy}
E_m&=&-(1-\frac{1}{2m^2\ell^2})\int d^2x{\sqrt{-\bar{g}}\over 32\pi
G}\{\bar{\nabla}^0h_m^{\mu\nu}\dot{h}_{m\mu\nu}
\},\\
E_M&=&(1-\frac{1}{2m^2\ell^2})\int d^2x{\sqrt{-\bar{g}}\over32\pi
G}\{\bar{\nabla}^0h_M^{\mu\nu}\dot{h}_{M\mu\nu} \}.\eea

By plugging in the solutions of highest weight gravitons
\cite{Li:2008dq} with our $(h,\bar h)$ into the above expression, we
find that for the massless modes $h_{\mu\nu}^m$, the energy is
positive for $m^2\ell^2>1/2$ and negative for $m^2\ell^2<1/2$; for
the massive modes $h_{\mu\nu}^M$, the energy is positive for
$m^2\ell^2<1/2$ and negative for $m^2\ell^2>1/2$. The same as in
chiral gravity, to have an asymptotically $AdS_3$ vacuum, we can
only have $m^2\ell^2=1/2$. However, at this point, the energies of
both highest weight massless and massive modes are zero, and both
the left moving and right moving central charges are zero. This can
be seen as a sign that the new massive gravity in asymptotically
$AdS_3$ may be trivial at this critical point and both the massless
and massive gravitons can be viewed as pure gauge in the framework
of Brown-Henneaux boundary conditions. If they are not pure gauge,
the $AdS_3$ vacuum at this critical point is not stable.

However, there are several points to note here. First, we have
assumed the Brown-Henneaux boundary conditions through the
calculations, but the consistency of these conditions with the new
massive gravity still need to be confirmed, as it has been done in
\cite{Hotta:2008yq} for topological massive gravity with a negative
cosmological constant. Also even if the consistency is satisfied we
still need to calculate the conserved charges to see if the charges
are all zero at the critical point. Second, even if the above
argument is satisfied, this does not mean that the new massive
gravity with a negative cosmological constant is trivial because we
may still have another stable vacuum which is not $AdS_3$ for other
values of the mass parameter, like the warped $AdS_3$ black holes in
topological massive gravity. Also there may be other ways to
calculate the energy of the gravitons which do not give this result.

\section{Conclusion and Discussion}

In this note, we studied the new massive gravity theory in
asymptotically $AdS_3$ spacetime, and found that this theory could
only be sensible at $m^2\ell^2=1/2$ in order to have an $AdS_3$
vacuum. However, at this special point, the energies of both the
highest weight massless and massive modes are zero, and both the
left moving and right moving central charges are zero. The new
massive gravity in asymptotically $AdS_3$ may be trivial at this
critical point in the sense of the Brown-Henneaux boundary
conditions. At this point, the boundary conditions of log gravity
\cite{{Grumiller:2008qz},{Grumiller:2008es},{Henneaux:2009pw},Maloney:2009ck}
may be preferred and give interesting results.

In the calculations we have assumed the Brown-Henneaux boundary
conditions, but we need to further analyze the consistency of the
boundary condition with the new massive gravity. Also we need to
check if the conserved charges are zero at the critical point. Even
if this theory is indeed trivial at the critical point we can have
another stable vacuum which is not locally $AdS_3$ for any value of
the mass parameter. And we can also relax the boundary conditions to
have more interesting physics in this theory like it was done in
\cite{{Grumiller:2008qz},{Grumiller:2008es},{Henneaux:2009pw},Maloney:2009ck}.

\section*{Acknowledgments}

We would like to thank Wei Song for very helpful and valuable
discussions. We would also like to thank Wei He for collaboration at
an early stage of this work. We are indebted to Ricardo Troncoso,
Daniel Grumiller and Niklas Johansson for useful comments and
enjoyable discussions on related subjects. This work was supported
in part by the Chinese Academy of Sciences with Grant No.
KJCX3-SYW-N2 and the NSFC with Grant No. 10821504 and No. 10525060.

\end{document}